\documentstyle[prl,aps,epsfig]{revtex}

\newcommand{\beq}{\begin{equation}}
\newcommand{\eeq}{\end{equation}}
\newcommand{\beqar}{\begin{eqnarray}}
\newcommand{\eeqar}{\end{eqnarray}}
\newcommand{\bfig}{\begin{figure}}
\newcommand{\efig}{\end{figure}}

\newcommand{\bd}{\begin{itemize}} 
\newcommand{\ed}{\end{itemize}} 
\newcommand{\bc}{\begin{center}}
\newcommand{\ec}{\end{center}}
\newcommand{\be}{\begin{equation}}
\newcommand{\ee}{\end{equation}}

\newcommand{\ba}{\begin{array}}
\newcommand{\ea}{\end{array}}

\newcommand{\set}[2]{\newcommand{#1}{#2}}
\set{\pa}{\partial \over \partial\, }
\set{\leftvector}{\stackrel{\leftarrow}{\partial }}
\set{\rightvector}{\stackrel{\rightarrow}{\partial }}
\begin{document}
\twocolumn[\hsize\textwidth\columnwidth\hsize
           \csname @twocolumnfalse\endcsname

\title{What about a beta-beam facility for low-energy neutrinos ?}
\author{Cristina Volpe$^{a,b}$}
\address{$^{a)}$
  Groupe de Physique Th\'{e}orique, Institut de Physique Nucl\'{e}aire,
F-91406 Orsay Cedex, France\\
$^{b)}$
Institut f\"ur Theoretische Physik der Universit\"at Heidelberg, Philosophenweg
19, D-69120 Heidelberg, Germany \\}

\maketitle

\begin{abstract}
A novel method to produce neutrino beams has recently been proposed :
the beta-beams. This method consists in using the $\beta$ decay of 
boosted radioactive
nuclei to obtain an intense, collimated and pure neutrino beam.
Here we propose to exploit the beta-beam concept to produce neutrino beams 
of low energy. We discuss the applications 
of such a facility as well as its
importance for different domains of physics. We focus, in particular, on
neutrino-nucleus interaction studies of interest for various open issues 
in astrophysics, nuclear and particle physics. We suggest
possible sites for a low-energy beta-beam facility.

\end{abstract}

\vskip2pc
]

\noindent
\newpage 
\noindent
The recent discovery that neutrinos are massive particles
has considerable impact on different domains of physics:
in particle physics, where the description of non-zero masses
and mixing requires the extension of the Standard Model of fundamental 
interactions; in astrophysics, for the comprehension of various
phenomena such as nucleosynthesis; in cosmology with, for 
instance, the search for dark matter.

In the last few years 
 positive oscillation signals have been found in a series of experiments
using neutrinos produced with various sources~\cite{sk,lsnd}. 
In view of the importance of this
discovery and its implications, a number of projects are 
running, planned in the near future, or under study
in order to address many
still open questions about neutrinos. 
Among them are those concerning their Majorana or Dirac
nature, the mass hierarchy and absolute mass scale, 
the knowledge of the mixing
angle $\theta_{13}$, the possible existence of sterile
neutrinos and of $CP$ violation in the leptonic sector.

In a recent paper~\cite{zucchelli}
Zucchelli has proposed an original method to
produce intense, collimated and pure neutrino beams: the {\it beta-beams}.
In contrast with the
neutrino factory concept implying the production, collection and storage of
muons to obtain muon and electron neutrino beams,
the novel method consists in accelerating, to high energy,
radioactive ions
decaying through a $\beta$  process.
A beta-beam facility consists of a radioactive ion production
and
acceleration to low energy (like at CERN ISOLDE),
further
acceleration to about 150 GeV/nucleon
(using for example the PS/SPS accelerators at CERN) and storing 
of the radioactive ion bunches in a storage ring.
At present
$^{6}He$ and $^{18}Ne$ seem to be  the best candidates~\cite{zucchelli}.
The resulting neutrino beam has three novel features, namely a single
neutrino flavor (electron neutrino or anti-neutrino), 
a well-known energy spectrum and intensity, 
a strong collimation. Another important advantage:
a beta-beam scheme relies on existing technology.
The physics impact of  
such a beam has been discussed in~\cite{zucchelli,mauro}
and includes for example oscillation searches, precision physics and 
$CP$ violation measurements.
The feasibility of beta-beams is at present under 
careful study~\cite{betabeam}.

In this letter we propose to exploit the beta-beam concept 
to produce intense, collimated and pure 
neutrino beams of low energies~\cite{cris}.
Low energies means here a few tens of MeV, like those 
involved in nucleosynthesis and in supernova explosions, up to
about a hundred MeV.
We argue that the physics potential of such a facility would have an
important 
impact on hot issues 
in different domains, in particular nuclear physics,
particle physics
and astrophysics. To illustrate this we focus on the specific example
of neutrino-nucleus interaction studies and discuss some open
questions that could be addressed with a low-energy beta-beam facility~\cite{cris}.
Finally, we analyze possible sites for such a facility.

Nuclei are used to detect neutrinos in 
experiments designed to study neutrino properties, such as
oscillation measurements, as well as experiments where neutrinos
bring information from
the interior of stars like
our sun or from supernova explosions.
As a consequence, 
a detailed understanding of neutrino induced reactions
on nuclei is crucial both for the
interpretation of various current experiments and  
for the evaluation of the feasibility and physics potential of 
new projects.
Examples are given by the use of~\cite{sk}: - deuteron in heavy water detectors
like in SNO for solar neutrinos; - carbon in scintillator detectors
such as in the LSND and KARMEN experiments using neutrinos from a beam 
dump~\cite{lsnd,karmen}; -
oxygen in Cherenkov detectors like in the Super-Kamiokande detector
or in next-generation large water detectors
like UNO and Hyper-K~\cite{uno}; - lead-perchlorate~\cite{elliott} 
and lead 
in new projects for supernova neutrinos such as
OMNIS and LAND~\cite{land}.
Open issues in astrophysics provide 
important motivations for improving our present knowledge of
neutrino-nucleus interactions. 
In particular, 
the role of these reactions for nucleosynthesis
is 
under intense investigation~\cite{qian}.  

So far, experimental data on neutrino-nucleus interactions
are extremely scarce.
The largest ensemble of data has been obtained for
carbon~\cite{lsnd,karmen}  
where discrepancies between experimental
and theoretical values have been the object of intensive studies
in the last years~\cite{cris1,c12}.
There is one measurement in deuteron~\cite{deut} 
and one in iron~\cite{iron}.
In the case of deuteron, where theoretical predictions are 
very accurate, there is still an important unknown
quantity, i.e. $L_{1A}$~\cite{kubodera}.
Theoretical calculations are therefore of absolute necessity.
However, getting accurate predictions is a challenging task
and necessitates as much experimental information as possible.

The general expression for the cross section
of the \mbox{$\nu_{l} + ^{A}_ZX_N \rightarrow l + ^{A}_{Z+1}X_{N-1}$} reaction
($l$ is the outgoing lepton), as a function of the incident neutrino energy $E_{\nu}$, 
  is given by~\cite{kuramoto} :
\begin{equation}
\sigma(E_{\nu})={G^{2} \over {2 \pi}}cos^{2}\theta_C\sum_{f}p_lE_l
\int_{-1}^{1}d(cos \, \theta)M_{\beta},
\label{e:1}
\end{equation}
where
$G \,cos \, \theta_C$ is the weak coupling constant, $\theta$ is
the angle between the directions of the incident neutrino and the outgoing
lepton, $E_l=E_{\nu}-E_{fi}$
is the outgoing lepton
energy  and $p_l$ its momentum, 
$E_{fi}$ being the energy transferred to the nucleus.
The quantity $M_{\beta}$ contains
the nuclear Gamow-Teller and Fermi type transition probabilities~\cite{cris1}.

The energy which can be transferred to the nucleus
in a neutrino-nucleus interaction does not have any upper value
since the neutrinos can have any impinging energy according
to the specific neutrino source. 
Typical neutrino energies 
cover the range
from the very low (up to about 10 MeV for reactor and solar neutrinos)
to the low (tens of MeV for e.g. supernova neutrinos)
energy regime,
 to the intermediate
(about 100-300 MeV) and high  
(GeV and multi-GeV) energy range of accelerator
and atmospheric neutrinos.
The nuclear degrees of freedom relevant in these various
energy windows
are very different and the models used to describe
the transition probabilities in (\ref{e:1}) 
range from the Elementary Particle Model, Effective Field Theories, 
detailed
microscopic approaches 
(Shell Model,
Random-Phase Approximation and its variants) for low momentum transfer,
to the Fermi Gas Model at high momentum transfer \cite{kuboreview}.

One of the difficulties in getting accurate theoretical predictions
comes from the increasing role played by the forbidden transitions
when the neutrino energy increases,
as pointed out in~\cite{cris1,cris2}.
The importance of the forbidden spin-dipole transitions
in nucleosynthesis has been first pointed out in \cite{gail1}.
As an example
Fig.1 shows the contribution 
of various states, excited in the
$\nu_e$(Pb,Bi)$e^-$ reaction, to the total cross section and its evolution
with increasing neutrino energy.
In particular we see that
already for 30-50 MeV neutrino energy
the contribution of forbidden states ($J^{\pi} \neq 0^+,1^+$)
becomes significant.

The importance of forbidden states can also be seen
directly in the flux-averaged cross sections
-- obtained by folding the cross sections (\ref{e:1})
with the relevant neutrino flux -- which
are the relevant quantities for experiments.
For low energy neutrino, such as supernova neutrinos, or 
neutrinos produced by the decay-at-rest of muons,
the spin-dipole states ($J^{\pi}=0^-,1^-,2^-$)
contribute by about $40 \%$ in $^{12}$C~\cite{cris1}
and $^{56}$Fe~\cite{iron}, and by about $68 \%$ in $^{208}$Pb~\cite{cris2}.
The contribution from higher forbidden states is about $5 \%$ and $25 \%$
in iron and lead respectively.
Their role increases with increasing neutrino
energy. Indeed,
they contribute
by about $30 \%$ in carbon~\cite{cris1} and $60 \%$ in lead~\cite{cris2}
in the intermediate energy region corresponding, for example,
to neutrinos produced from pion decay-in-flight.

Few data exists on the spin-dipole states, mainly from charge-exchange 
reactions \cite{pn}
and practically none for the higher forbidden states\footnote{Some knowledge about the relevant states can be obtained 
through muon capture experiments.}. 
More experimental information is needed to constrain theoretical
calculations of the centroid, the width and
the total strength of forbidden states. 
For example,
one of the open questions concerning these states
is the possible quenching of their strength.
Note that 
understanding the quenching of the allowed Gamow-Teller ($J^{\pi}=1^+$)
strength,  namely the reason why the observed
strength is only a fraction of the predicted one, has
been a longstanding problem in nuclear physics~\cite{gt}.

This
has a direct impact on the physics potential of
running experiments or projects under study.
Let us consider the case of lead-based projects which aim at measuring
supernova neutrinos. It has been shown, for example,
that the a precise measurement of the energy of the electrons
emitted in the charged-current neutrino-lead reaction can
provide useful information about the temperature of the initial
muon/tau neutrinos  produced in a supernova explosion~\cite{cris3}.
Although this result seems little sensitive to the details
of the calculations, a measurement of the differential electron cross section
would bring an important piece of information.
Moreover, the number of charged current events in coincidence with
neutrons produced in the des-excitation of $Bi$
may be used to determine whether the mixing angle $\theta_{13}$
is much larger or much smaller than $10^{-3}$.
In the latter case, one would need -- 
as far as the neutrino detection is
concerned -- a very precise knowledge of the reaction cross 
sections~\cite{cris3}.
Similar studies have been performed in various other nuclei.
In \cite{petr} it has been shown that in Cherenkov detectors
the detection of $\gamma$ rays produced
in the inelastic neutrino scattering off
oxygen allow to identify $\nu_{\mu,\tau}$.

A low-energy beta-beam facility would provide the possibility to
perform neutrino-nucleus interaction studies with various nuclei
and address the many open questions \cite{kuboreview,orland,baha1}.
Examples are the measurements of reaction cross sections on deuterium,
carbon, oxygen, iron and lead.
In the latter case, 
the measurement of the differential electron cross section
as well as of the neutral and charged current cross sections
in coincidence with one- and two-neutron emission would be of
great interest.
A larger set of experimental data would allow us to make reliable
extrapolation from the low to the high neutrino energy regime.
It would also provide important information 
for the extrapolation to the case of neutrino reaction 
on exotic nuclei, which are of astrophysical 
interest.
Finally, one should
reanalyze, in the context of a low-energy beta-beam facility,
the feasibility of the experiments proposed for the ORLAND project
(Oak Ridge Laboratory for Neutrino Detectors)~\cite{orland}
which has been proposed a few years ago (these include, for example, 
oscillation searches, measurement of 
the Weinberg angle at low momentum transfer).
Another aspect of beta-beams should be stressed :
the neutrons emitted from some beta-decay candidates 
also open other axes of research besides the
one mentioned here. 

The future availability of intense radioactive ion
beams  at several facilities 
offers various possible sites for a beta-beam facility producing
low-energy neutrinos. Among these are GANIL, GSI, CERN or the EURISOL
project. Table 1 shows the capabilities (energy and intensities) which
can be attained at these sites. Concerning GSI, 
lower intensities will be reached with the presently
envisaged upgrade \cite{gsi}.  
We see that two configurations are possible. In sites like GANIL and
for the EURISOL project (in the present shape where the ions are
accelerated up to a 100 MeV/A and without a storage ring), the gamma
of the parent ions is equal to one. Therefore, one can bring the ions
in a $4\pi$ detector and dispose of intense neutrino sources.
In sites like GSI and CERN, the ions will be accelerated and stored in
a storage ring (at GSI with the future HESR). 
In particular, at GSI one will dispose of neutrinos
spanning the tens of MeV energy range, whereas at CERN, one could
span from tens to 100 MeV neutrino energy domain.

In conclusion, we propose to exploit the beta-beam concept to
produce intense and pure low energy neutrino beams.
Such a facility would have a considerable impact 
in different domains of physics.
Possible sites include CERN, GSI and GANIL.

\vspace{.5cm}
\noindent
I am grateful to M. Lindroos 
for the discussions concerning the feasibility of a low-energy
beta-beam facility as well as for suggesting the GANIL and GSI laboratories
as possible sites. 
Thanks also to A. Villari and H. Weick for fruitful discussions and 
to R. Lombard and J. Serreau for careful reading of
this manuscript.

\newpage

\begin{figure}
\begin{center}
\includegraphics[angle=-90.,scale=0.4]{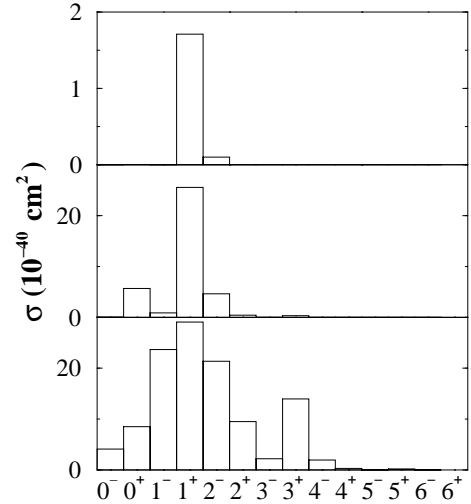}
\end{center}
\protect\caption{Contribution of states of different multipolarities to
the $^{208}$Pb$(\nu_{e},e^-)^{208}$Bi reaction cross section $(10^{-40}~$cm$^2)$
for $E_{\nu_e}=15~$MeV (up), $30~$MeV (middle),
$50~$MeV (bottom) \protect\cite{cris2}.}
\end{figure}

\begin{table}
\begin{tabular}{lcc} 
 & Ion intensity & $\gamma$   \\ \hline
GANIL & $10^{12}~$ions/s & 1 \\
EURISOL & $10^{13}~$ions/s & 1   \\
CERN & $2 \times 10^{13}$ions/s & 1-150 \\ 
\end{tabular}
\protect\caption{
The table shows the ion intensities and the gamma of the parent ion
which could be available 
at possible sites for a low-energy beta-beam facility.
The numbers refer to $^{6}$He as an example.
Results for CERN are from \protect\cite{betabeam}.}
\end{table}
\end{document}